\documentclass[twocolumn,twoside,letterpaper,11pt]{article}
\usepackage{graphicx,xrrc,natbib}
\usepackage{times}

\begin{document}


\title{OBSERVATIONS OF JET DISSIPATION}


%
%
%
%


\author{    R. A. Laing                                  } 
\institute{ European Southern Observatory                } 
\address{   Karl-Schwarzschld-Stra\ss e 2, D-85748 Garching-bei-M\"{u}nchen,
  Germany  } 
\email{     rlaing@eso.org                               } 

\author{    J. R. Canvin, A. H. Bridle                      }
\email{     jrc@astro.ox.ac.uk, abridle@nrao.edu                     }


\maketitle

\abstract{We summarise the recent progress of a project to determine 
  physical conditions in the jets of low-luminosity (FR\,I) radio galaxies. We
  model the jets as intrinsically symmetrical, relativistic, decelerating
  flows. By fitting to deep, high-resolution VLA images in total intensity and
  linear polarization, we have been able to derive for the first time the
  three-dimensional distributions of velocity, rest-frame emissivity and
  magnetic-field structure. We describe our results for four sources and outline
  the implications for theories of jet deceleration. We also summarise a
  conservation-law analysis of 3C\,31, which gives profiles of internal
  pressure, density, Mach number and entrainment rate along the jets, again for
  the first time. Finally, we discuss the applicability of adiabatic models to
  FR\,I jets and identify the regions where distributed particle acceleration
  is necessary.}

\section{Introduction}

It was first recognised by \cite{FR74} that the ways in which extragalactic jets
dissipate energy to produce observable radiation differ for high- and
low-luminosity  sources. Jets in weak FR\,I sources are bright close to the nucleus
of the parent galaxy, whereas those in powerful FR\,II sources are relatively
faint until their terminal hot-spots [there is also a dependence on environment
\citep{LO}]. It rapidly became accepted that FR\,I jets must decelerate by
entrainment of the surrounding IGM
\citep{Baa80,Beg82,Bic84,Bic86,DeY96,DeY04,RHCJ99,RH00} or by injection of mass lost
by stars within the jet volume \citep{Phi83,Kom94,BLK96}.

More recently, evidence has accumulated that FR\,I jets are initially
relativistic and decelerate on kpc scales. FR\,I sources are thought to be the
side-on counterparts of BL Lac objects, in which relativistic motion on parsec
scales is well-established \citep{UP95}.  Superluminal motions have been seen on
milliarcsecond scales in several FR\,I jets \citep{Giovannini01} and on
arcsecond scales in M\,87 \citep{Biretta95}. In FR\,I sources, the lobe
containing the main (brighter) jet is less depolarized than the counter-jet lobe
\citep{Morganti97}. This can be explained if the main jet points toward the
observer, suggesting that the brightness asymmetry is caused by Doppler beaming
\citep{Laing88}. The asymmetry decreases with distance from the nucleus
\citep{LPdRF}, consistent with the idea that the jets decelerate.

It has proved frustratingly difficult to quantify the basic properties --
composition, velocity, density, pressure, magnetic field -- of radio jets.  The
present paper summarises recent progress on a project which has, for the first
time, allowed us to derive the three-dimensional distributions of physical
parameters in FR\,I jets on kpc scales. There are three main elements:
\begin{enumerate}
\item We model FR\,I jets as intrinsically symmetrical, axisymmetric,
  relativistic flows. We parameterise the geometry and the three-dimensional
  variations of velocity, emissivity and magnetic-field structure and compute
  the brightness distributions in total intensity and linear polarization. By
  fitting to radio images, we can optimise the model parameters. We refer to
  these fits as {\em free models}. They are empirical, and make as few
  assumptions as possible about the (poorly-known) internal physics of the
  jets. Deep,
  high-resolution radio images in Stokes $I$, $Q$ and $U$ are essential.
\item We then apply conservation of mass, momentum and energy to infer the
  variations of pressure, density, entrainment rate and Mach number along the
  jets. This requires measurements of the external density and pressure profiles
  from X-ray observations.
\item Finally, we attempt to model the acceleration and energy-loss
  processes, adding images at mm, IR, optical and X-ray wavelengths to our radio
  data and models. So far, we have concentrated on the simplest process --
  adiabatic energy loss -- in an attempt to quantify the amount and location of
  distributed particle acceleration.
\end{enumerate}
Two X-ray -- radio connections are vital to our argument. Firstly, the bases
of FR\,I jets emit X-ray synchrotron radiation, indicating continuing
local particle acceleration. Secondly, the jets interact with and are confined by
the hot intergalactic medium, whose density and temperature can be estimated
from its thermal bremsstrahlung emission. The radio data alone provide
constraints on the {\em kinematics} of the jets.  Combining these kinematics
with the densities and pressure gradients obtained from the X-ray data
via a conservation-law analysis allows us to address the jet {\em dynamics}
and to estimate Mach numbers, energy and mass fluxes and entrainment rates 
in the jets.
 
We began this project by making a statistical study of jet asymmetries in the B2
sample \citep{LPdRF}. The detailed models were first applied to the well-known
FR\,I source 3C\,31 by \cite{LB02a} and a conservation-law analysis of the same
source was made by \cite{LB02b}, using {\em Chandra} observations by
\cite{Hard02}. Adiabatic models were developed and applied, again to 3C\,31, by
\cite{LB04}. More recently, the techniques have been extended to other sources
and free models of 0326+39 and 1553+24 were published by \cite{CL04}.
In the present paper, we summarise the published results and also present models
of the jets in NGC\,315. Modelling of one further source, 3C\,296
\citep{Hard97}, is in progress.

\section{Free models}
\label{Free}

\subsection{Model procedure}

Our fundamental assumption is that the jets are intrinsically symmetrical,
axisymmetric, relativistic, stationary flows. [The assumption of symmetry is
critical to our approach and we consider it further in
Section~\ref{symmetry}]. The magnetic fields are assumed to be disordered on
small scales, but anisotropic. There are both observational and theoretical
reasons to suppose that the longitudinal field component, at least, cannot be
vector-ordered on large scales \citep{BBR,Laing81,LB02a}. We return to the
question of the nature of the toroidal field component in Section~\ref{Field}.

The main elements of the modelling procedure are as follows:
\begin{enumerate}
\item We parameterise the geometry, velocity, emissivity and field structure
  using simple, analytical expressions \citep{LB02a,CL04}.
\item For given model parameters, we derive model distributions of Stokes $IQU$
  by integration along the line of sight, taking account of the anisotropy of
  synchrotron radiation in the rest frame of the emitting material, aberration
  and Doppler beaming \citep{Laing02}.
\item The model is then compared with deep VLA images, using $\chi^2$ (summed
  over Stokes parameters) as a measure of goodness of fit and the parameters are
  optimised using the downhill simplex method of \cite{NL65}.  We take a measure
  of the deviation of the brightness distributions from axisymmetry as an
  estimate of the ``noise'' (rather than the much smaller off-source rms) and
  obtain reduced $\chi^2$ values in the range 1.1 -- 1.7 over 1200 - 1400 
  independent points, each with 3 Stokes parameters.  
\end{enumerate} 

\subsection{The assumption of intrinsic symmetry}
\label{symmetry}
It is, of course, an assumption that the jets in most FR\,I galaxies are
intrinsically symmetrical on small scales -- and one that is difficult to prove
in any individual case [there are also some clear exceptions, such as 0755+37
\citep{Bondi00}]. It is also obvious that many jets are very far from symmetry
on large scales. We contend, however, that the apparent asymmetries caused by
aberration and Doppler beaming dominate over intrinsic and environmental
differences close to the nuclei in most sources. Evidence in favour of this
hypothesis includes:
\begin{enumerate}
\item The intensity ratio between main and counter jets initially decreases with
distance from the nucleus \citep{LPdRF}, reaching a value close to unity on
scales $\sim$10\,kpc in typical cases. Deceleration provides a natural mechanism
to reduce the asymmetry if the flow is relativistic, whereas intrinsic or
environmental differences between the jets would be likely to produce constant
or increasing brightness differences. The fact that the jets are very
symmetrical on intermediate scales indicates that intrinsic differences are
small.
\item Jet/counter-jet intensity ratios are well correlated with other indicators
  of relativistic motion, such as the core/extended flux ratio \citep{LPdRF}.
\item In those cases where superluminal motion is observed, it is always on the
  side of the brighter kpc-scale jet \citep{Biretta95,Giovannini01}.
\item The observation that the lobe containing the brighter kpc-scale jet is
  less depolarized than the counter-jet lobe in FR\,I sources \citep{Morganti97}
  indicates that the brighter jet is seen through less magnetoionic material and
  is therefore on the near side, as expected.
\item There is a characteristic asymmetry in the degree and direction of
  polarization (corrected for Faraday rotation), which is correlated with that
  in total intensity. We show below that relativistic aberration provides a
  natural explanation for this effect.
\end{enumerate}

\begin{figure*}
  \begin{center}
    \includegraphics[width=16cm]{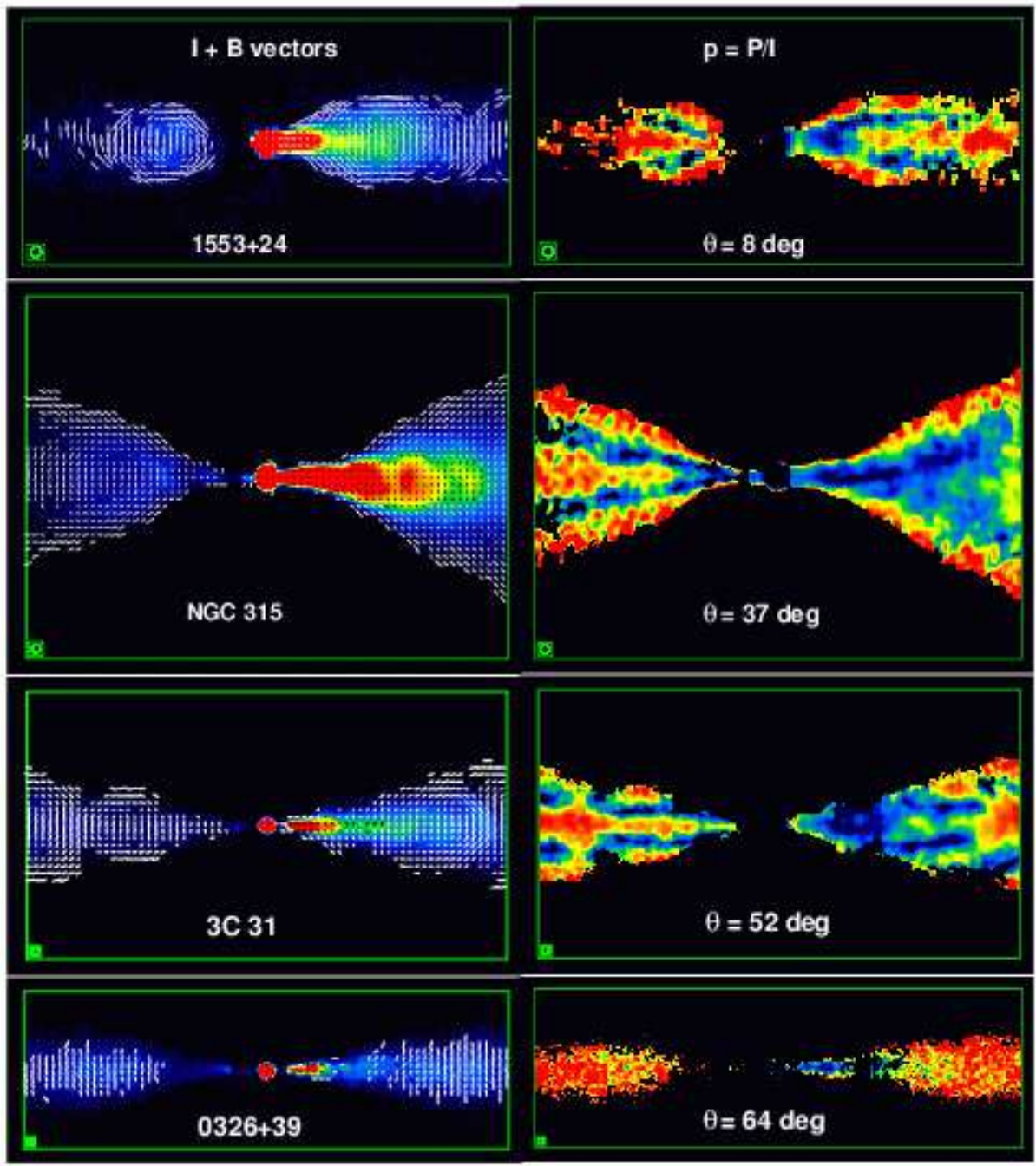}
    \caption{The observed total-intensity and polarization structures of the
    four modelled sources. The sources are arranged from top to bottom in order
    of increasing angle to the line of sight, $\theta$, as deduced from the
    model fits. The left-hand panels show vectors with lengths  proportional
    to the degree of polarization, $p$ and directions along the apparent
    magnetic field superimposed on false-colour plots of total intensity. The
    right-hand panels show false-colour plots of $p$, in the range 0 (blue) --
    0.7 (red). Note: (a) the asymmetry in total intensity decreases with
    increasing $\theta$; (b) there is a characteristic difference between the
    polarization structures of the main and counter-jets (see text) and (c)
    0326+39 has no parallel-field edge, unlike the other three sources. Data
    from \cite{CL04} (0326+39 and 1553+24), \cite{LB02a} (3C\,31) and Cotton et
    al., in preparation (NGC\,315).\label{fig:piv}}
  \end{center}
\end{figure*}

Fig.~\ref{fig:piv} illustrates the characteristic asymmetries in total
intensity and linear polarization for the four sources in our study. They are
arranged in order of increasing angle to the line of sight, $\theta$ (as deduced
from the model fits).  The asymmetries in total intensity decrease with distance
from the nucleus and are, of course, greater at smaller $\theta$. In addition,
there is an asymmetry in the polarization structure: the main jet base shows a
polarization minimum on-axis which extends considerably further from the nucleus
than any equivalent in the counter-jet. Our (intrinsically symmetrical) models
successfully reproduce this difference (see Section~\ref{comparison}) and we
believe that it, like the intensity difference, is a signature of relativistic
aberration.

\subsection{Breaking the degeneracy between velocity and angle}

The key to our modelling approach is its use of asymmetries in linear
polarization as well as total intensity in order to break the degeneracy between
velocity and angle.  The standard formula for the ratio of jet and counter-jet
flux densities (assuming isotropic emission in the rest frame) is:
\begin{eqnarray*}
  I_{\rm j} / I_{\rm cj} & = & \left (\frac{1 + \beta\cos\theta}{1 -
  \beta\cos\theta} \right )^{2+\alpha} \\
\label{eq:iratio}
\end{eqnarray*}
where $\alpha$ is the spectral index and $\beta = v/c$. There is therefore a
degeneracy between velocity and angle to the line of sight.  The relation
between the angles to the line of sight in the rest frame of the flow,
$\theta^\prime$ and in the observed frame, $\theta$, is:
\begin{eqnarray*} 
\sin\theta^\prime_{\rm j} & = & [\Gamma(1-\beta\cos\theta)]^{-1}\sin\theta
\makebox{~~~~~(main jet)} \\
\sin\theta^\prime_{\rm cj} & = & [\Gamma(1+\beta\cos\theta)]^{-1}\sin\theta
\makebox{~~(counter-jet)} \\
\end{eqnarray*}
where $\Gamma = (1-\beta^2)^{-1/2}$ is the Lorentz factor. The observed
polarization is in general a function of $\theta^\prime$ if the field is
anisotropic.  If we know the field structure a priori, then we can solve
explicitly for $\beta$ and $\theta$. We take the example of a field which is
disordered on small scales but confined to a plane perpendicular to the jet,
with equal rms along any direction in the plane. In this case, there is no
variation of the degree or direction of polarization across the jet. For $\alpha
= 1$, the total and polarized flux densities per unit length in the emitted
frame are:
\begin{eqnarray*}
I^\prime & = & K(1+\cos^2\theta^\prime) \\
P^\prime & = & p_0 K\sin^2\theta^\prime \\
\end{eqnarray*}
where $K$ is a constant \citep{Laing80,Laing81}.  The ratios of
observed total and polarized intensity for the jet and counter-jet
are:
\begin{eqnarray*}
\frac{I_{\rm j}}{I_{\rm cj}} & = & \left[ \frac{2 -
    [\Gamma(1-\beta\cos\theta)]^{-2} \sin^2\theta}{2 -
    [\Gamma(1+\beta\cos\theta)]^{-2} \sin^2\theta} \right]^3
    \\
\frac{P_{\rm j}}{P_{\rm cj}} & = &
    \left(\frac{1+\beta\cos\theta}{1-\beta\cos\theta}\right)^5
    \\
\end{eqnarray*}
These equations can be solved numerically for $\beta$ and $\theta$
(cf. \citealt{Bondi00}).  In general, we must fit the field configuration and
therefore need to introduce additional parameters to describe its
anisotropy. These also determine the variation of polarization transverse to the
jet axis, and can be estimated independently of the velocity and angle if the
jets are well resolved in this direction.

The selection of sources to be modelled is determined by the requirement for
high signal/noise ratio and good transverse resolution in $I$, $Q$ and $U$ in
both jets. It is also essential to correct for Faraday rotation. 

\subsection{Comparison of models and data}
\label{comparison}

\begin{figure}
  \begin{center}
    \includegraphics[width=7.2cm]{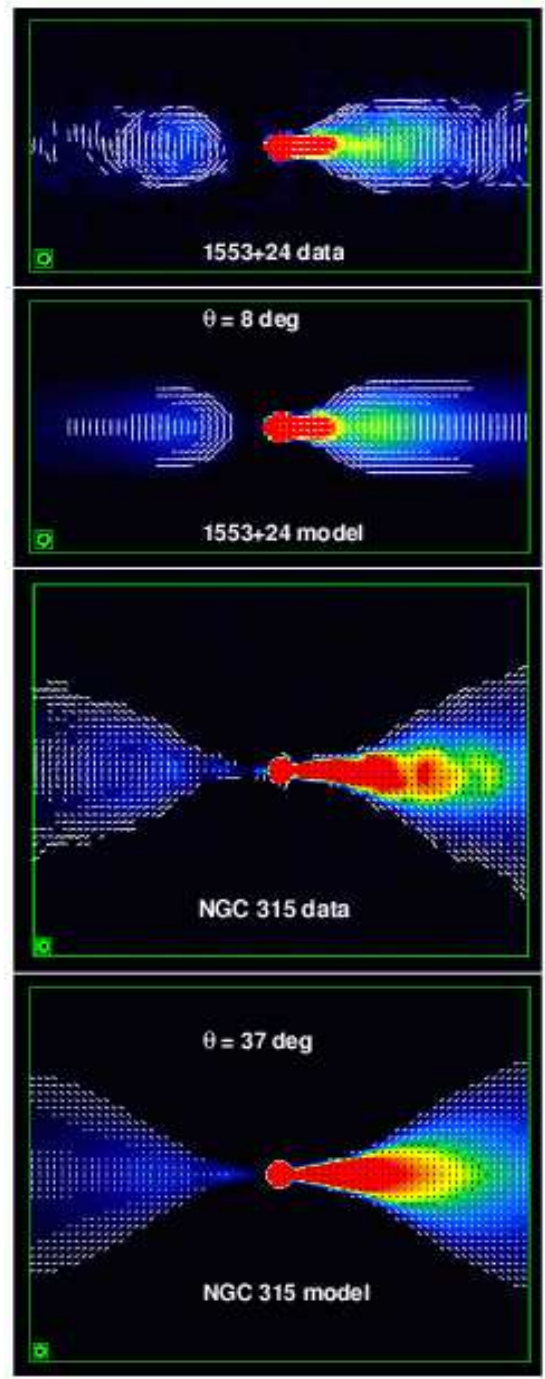}
    \caption{Comparison between the model and observed brightness distributions
    for 1553+24 (upper two panels) and NGC\,315 (bottom two panels). Vectors
    with lengths proportional to the degree of polarization and directions along
    the apparent magnetic field are superimposed on
    false-colour plots of total intensity. The resolutions are 0.75 and
    2.35\,arcsec FWHM, respectively. Data and models from \cite{CL04}
    (1553+24) and Cotton et al., in preparation (NGC\,315). \label{fig:compare1}}
  \end{center}
\end{figure}

\begin{figure}
  \begin{center}
    \includegraphics[width=\columnwidth]{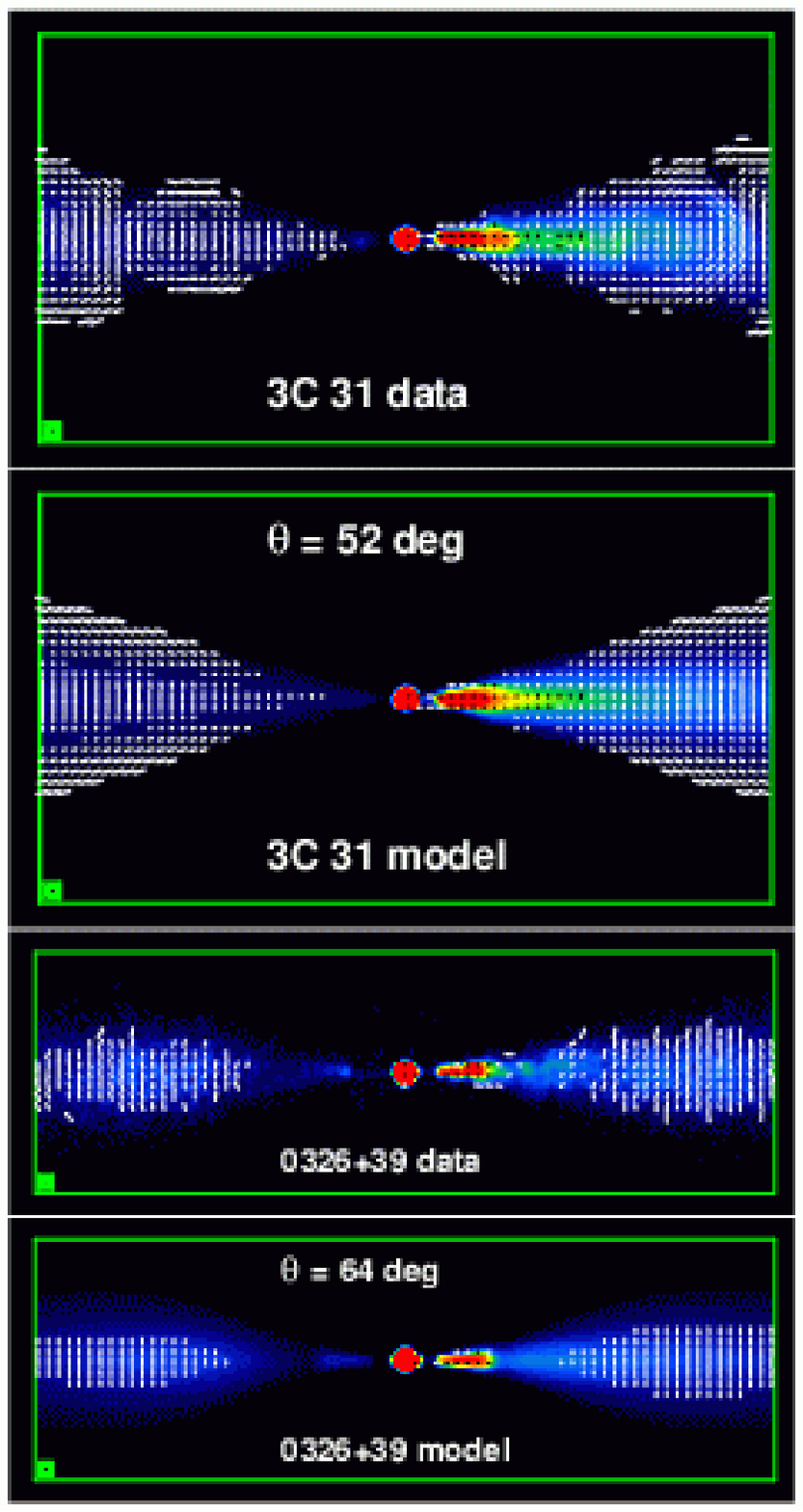}
    \caption{Comparison between the model and observed brightness distributions
    for 3C\,31 (upper two panels) and 0326+39 (bottom two panels). Vectors with
    lengths proportional to the degree of polarization and directions along the
    apparent magnetic field are superimposed on false-colour plots of total
    intensity. The resolutions are 0.75 and 0.5\,arcsec FWHM, respectively. Data
    and models from \cite{LB02a} (3C\.31) and \cite{CL04}
    (0326+39). \label{fig:compare2}}
  \end{center}
\end{figure}

\begin{figure}
  \begin{center}
    \includegraphics[width=6cm]{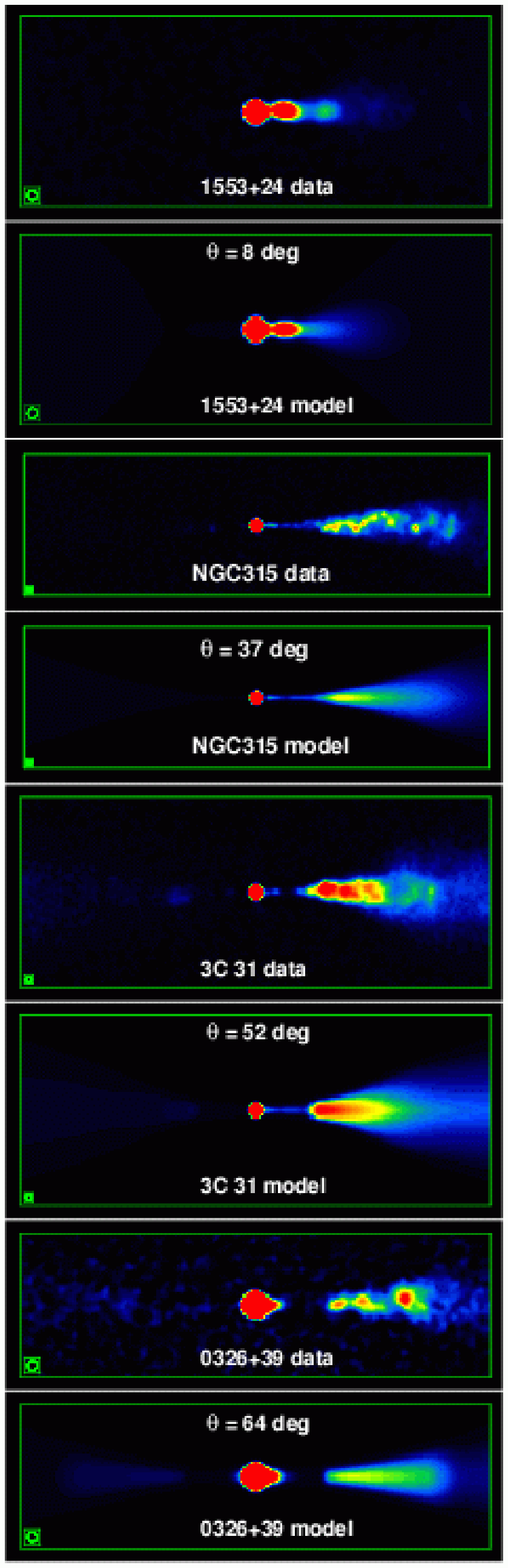}
    \caption{Comparison between model and observed total-intensity images at
    high resolution. Note the non-axisymmetric and knotty structure observed
    where the jets first brighten significantly. This cannot be described in
    detail by our axisymmetric models, but the average structure is reproduced
    correctly. NGC\,315 and 3C\,31 clearly show highly-collimated inner jets:
    these are probably present in the other two sources, but are too faint or
    small to image. References as in Figs~\ref{fig:compare1} and
    \ref{fig:compare2}.\label{fig:compareh}}
  \end{center}
\end{figure}

The four sources we have modelled are 3C\,31 \citep{LB02a}, 0326+39, 1553+24
\citep{CL04} and NGC\,315 (Cotton et al., in preparation). In
Figs~\ref{fig:compare1} -- \ref{fig:compareh} we compare model and observed
images at the same resolution. Figs~\ref{fig:compare1} and \ref{fig:compare2}
show the full modelled areas at moderate resolution and display both total
intensity and linear polarization. They demonstrate that we can successfully fit the
observations in all cases, limited primarily by small-scale (and often
non-axisymmetric) fluctuations. The inferred angles to the line of sight are
shown in the figures.  Fig.~\ref{fig:compareh} shows the inner parts of the
model and observed $I$ images at higher resolution. Here, the models fit the
average intensities very well, but cannot reproduce the complex small-scale
structure. 

\section{Physical parameters deduced from the free models}
\label{Physparms}

\subsection{Geometry}

Our qualitative conclusions about the geometry of FR\,I jets are not new, but
our estimation of angle to the line of sight allows us to deproject their shapes
and to derive intrinsic parameters. FR\,I jets are initially narrow, expand
abruptly and then recollimate. We refer to this part of the flow as the {\em
flaring region}. The shape of its outer isophote can be fitted accurately by a
cubic polynomial function of distance along the axis. Further from the nucleus,
the {\em outer region} is conical. In all sources except 3C\,31 the opening
angle is very small, so that the outer regions are almost cylindrical (note that
we model only the flaring region in NGC\,315). The deprojected shapes of the
sources are shown in Fig.~\ref{fig:vels}. The very small angle to the line of
sight inferred for 1553+24 means that its intrinsic shape is very narrow (see
Section~\ref{unified}).

\subsection{Velocity}

\begin{figure}
  \begin{center}
    \includegraphics[width=\columnwidth]{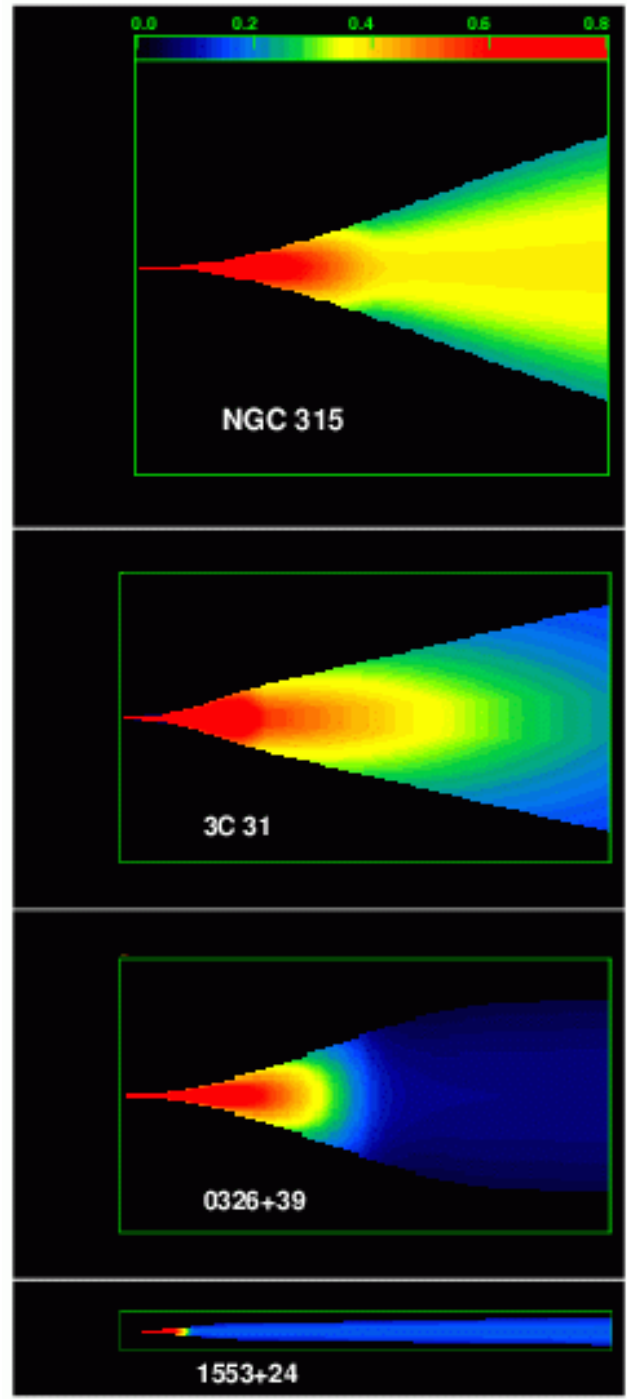}
    \caption{The velocity structures deduced for the four modelled sources. The
    labelled wedge indicates values of $\beta$ in the range 0 -- 0.8.  All of
    the jets have $\beta \approx 0.8$ where we first resolve them and decelerate
    rapidly over a short distance in the flaring region.  Further out, the
    velocities are either roughly constant (NGC\,315, 0326+39 and 1553+24) or
    decrease more slowly (3C\,31).\label{fig:vels}}
  \end{center}
\end{figure}

Fig.~\ref{fig:vels} also shows the inferred velocity distributions.  At the
closest distance from the nucleus where we can model the jets, all of their
on-axis velocities are close to $\beta \approx 0.8$. They all decelerate rapidly
over a short distance in the flaring region, but at different distances from the
nucleus. NGC\,315, 0326+39 and 1553+24 then have roughly constant velocities in
the range $\beta \approx$ 0.1 -- 0.4 until the end of the modelled
regions. 3C\,31 continues to decelerate, but more slowly.

There is clear evidence for transverse velocity gradients in 3C\,31 and
NGC\,315.  In the other two sources, the gradients are poorly determined, but in
all four cases, the data are everywhere consistent with a constant ratio of
edge/on-axis velocity $\approx 0.7$. There are no obvious low-velocity wings.

\subsection{Emissivity}

\begin{figure}
  \begin{center}
    \includegraphics[width=\columnwidth]{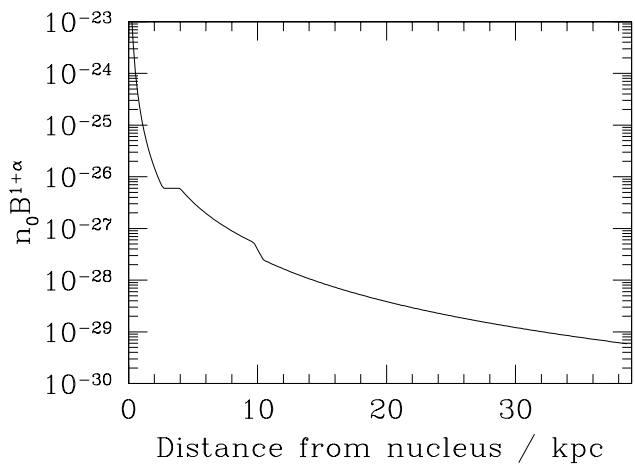}
    \caption{A profile of $n_0 B^{1+\alpha}$ along the jet axis for the model of
    NGC\,315, deduced from the emissivity. The energy spectrum of the radiating
    electrons is $n_0 E^{-2\alpha+1} dE$ and $B$ is the magnetic field. $n_0$ is
    in m$^{-3}$ and $B$ in T. Note the region of enhanced emissivity from 2.5 to
    10\,kpc from the nucleus. This corresponds to the knotty, quasi-helical
    structure in the main jet (Fig.~\ref{fig:compareh}).\label{fig:ngc315emiss}}
  \end{center}
\end{figure}

The longitudinal emissivity profiles generally flatten with increasing distance
from the nucleus. All of the sources show an anomalously bright region, with
complex, knotty sub-structure, within the flaring region
(e.g. Fig.~\ref{fig:ngc315emiss}).  This has an abrupt start, corresponding to
the {\em flaring point} of the jet \citep{Parma87,LPdRF} and often also a clear 
end: it is as if an extra component of emissivity is added to an underlying,
smooth distribution. \cite{LB02a} suggested that the sudden brightening
coincides with a change in jet collimation properties in 3C\,31, but the
transverse resolution is marginal for that source, and there appears to be no
obvious change in collimation in NGC\,315, the only well-resolved case. 

The jets are all intrinsically centre-brightened, with a typical edge/centre
emissivity ratio $\approx 0.2$.

\subsection{Field structure}
\label{Field}

We have, for the first time, estimated the three-dimensional field structures of
the jets rather than their projections on the plane of the sky.  The dominant
field component at large distances from the nucleus is {\em toroidal}. The
longitudinal component can be significant close to the nucleus, but decreases
further out, sometimes becoming negligible. The behaviour of the radial
component is peculiar, and varies from source to source.  This field evolution
is qualitatively consistent with flux freezing, but laminar-flow models, even
including transverse velocity shear, do not fit quantitatively (see
Section~\ref{Adiabatic}).

As mentioned earlier, the idea that the longitudinal and toroidal field
components are both vector-ordered (forming something like a simple helix) is
inconsistent both with the observation that the brightness and polarization
distributions are very symmetrical under reflection through their axes and with
constraints from flux conservation. The former argument is strongest on large
scales; weaker where the jets first brighten and there is significant oblique
structure. The hypothesis that the longitudinal component has many reversals,
but the toroidal component is well-ordered is consistent with the observations,
however (our calculations would give identical results in this case). It is
difficult to understand how a jet could acquire a toroidal component with many
reversals.

We have searched for the characteristic signature of Faraday rotation by a
toroidal field in the jets of 3C\,31 and NGC\,315, but see only random
fluctuations which are most plausibly associated with foreground galaxy or
group-scale plasma containing a disordered field. In any case, if our estimates
of internal jet density (Section~\ref{Conservation}) are correct, we would expect
negligible internal Faraday rotation even from a perfectly-ordered field.

Although the toroidal field component is probably not required for the
collimation of FR\,I jets in the regions we model (see below), it may be what
remains of a dynamically-important collimating field on smaller scales.

\subsection{Deceleration physics}

Any theory of jet deceleration has to explain the following results:
\begin{enumerate}
\item Jets have (at least) two regions differentiated by their collimation
  properties: flaring and outer.
\item The onset of rapid deceleration occurs within the flaring region and is quite
  sudden.
\item After the rapid deceleration, three of the sources (0326+39, 1553+24 and
 NGC\,315) have roughly constant velocities. 3C\,31 continues to decelerate, but
 more slowly. The difference may be connected to the opening angle of the
 conical outer region: 3C\,31 expands much more rapidly than the other three
 sources.
\item The sudden brightening of the jets occurs within the flaring region and
  significantly before the rapid deceleration. It is a consequence either of a
  real increase in rest-frame emissivity or of an abrupt slowing of the
  emissivity decline. Complex and clearly non-axisymmetric structures produce the
  excess emissivity.
\item This region of enhanced emissivity persists roughly until the start of
  rapid deceleration. Thereafter, the jets appear somewhat smoother and the
  emissivity declines monotonically.
\item There is evidence from the field structure of 3C\,31 for interaction with
  the external medium at the edge of the flaring region \citep{LB02a}, but it is
  not clear whether this effect is general.
\end{enumerate}
The reason for the sudden brightening of the jets is unclear, but
\cite{LB02a,LB02b} favour a reconfinement shock \citep{Sand83}, as this provides
a natural mechanism both for a sudden brightening at the same distance from the
nucleus in both jets and for the inferred over-pressure in 3C\,31 (see
Section~\ref{Conservation}). One interesting possibility is that entrainment of
the external medium, in the form of relatively large clumps,  begins at the
reconfinement shock, but that deceleration does not become significant until the
entrained material has time to mix with the jet flow.

\subsection{Implications for unified models}
\label{unified}

\subsubsection{Total intensity}

\cite{LB02a} show the appearance of the 3C\,31 model over a range of angles to
the line of sight, and demonstrate that it would be qualitatively consistent
with the appearance of the extended radio structures of BL Lac objects for small
$\theta$. 

Conversely, we find that 1553+24 has a very small value of $\theta \approx
8^\circ$. This means that its side-on counterparts must
be very large. The length of the main jet is at least 60\,arcsec
corresponding to 388\,arcsec (340\,kpc) at
$\theta = 60^\circ$, the median angle to the line of sight. This is comparable
in size with the longest jet in the B2 sample, in NGC\,315 \citep{Willis81}, and
far larger than the median ($\approx$30\,kpc; \citealt{Parma87}). There is cause
for suspicion unless: (a) the side-on counterparts of B2\,1553+24 are not
members of the B2 sample or (b) sources in that sample have linear sizes far
larger than previously realised.

\cite{CL04} examined the potential selection effects in detail and concluded
that the side-on counterparts of 1553+24 could have escaped identification,
either because they were missed in the original survey or because their angular
sizes have been greatly underestimated. This is because there is relatively
little lobe emission in 1553+24, so even its low-frequency flux is significantly
affected by Doppler beaming and because its outer jets would be extremely faint
and difficult to detect in published images.  Morganti \& Parma (private
communication) and Ledlow \& Owen (2004; in preparation) have made more
sensitive observations of radio galaxies in the B2 sample using the WSRT and the
VLA in D configuration and have shown that a significant fraction of them have
much longer radio jets than have previously been reported, extending many 100's
of kpc or even further. We suggest that these might include the missing
counterparts of 1553+24.

\subsubsection{Polarization}

Our models also require that the observed polarization structure changes as a
function of angle to the line of sight and predictions for the 3C\,31 model are
given by \cite{LB02a}.  The dominance of toroidal field at large distances from
the nucleus means that a parallel-field edge is less likely to be observed when
$\theta \approx 90^\circ$ \citep{Laing81}. Although the details of this effect
depend on the relative amounts of radial and longitudinal field present (these
differ between sources), it is interesting that 3C\,449 \citep{Fer99} and
PKS\,1333$-$33 \citep{KBE86}, which are extremely symmetrical and therefore
likely to be close to the plane of the sky, show no parallel-field edge; neither
does 0326+39 (which has the largest value of $\theta$ of any of our modelled
sources).

\section{Conservation-law analysis}
\label{Conservation}

\subsection{Assumptions and results for 3C\,31}

The free models described in the previous section tell us the velocity and
cross-sectional area of the jets. We can derive the external density and
pressure from X-ray observations, and the combination allows us to apply
conservation of matter, momentum and energy.  For 3C\,31, \cite{LB02b} showed
that well-constrained solutions exist subject to two key assumptions: 
\begin{enumerate}
\item $\Phi =
\Pi c$, where $\Phi$ is the energy flux (with rest-mass energy subtracted) and
$\Pi$ is the momentum flux.  This must hold quite accurately if the jets have
decelerated from bulk Lorentz factors $\approx 5$ on pc scales. 
\item The jets are in pressure balance with the external medium in their outer
  regions.
\end{enumerate}

\begin{figure}
  \begin{center}
    \includegraphics[width=\columnwidth]{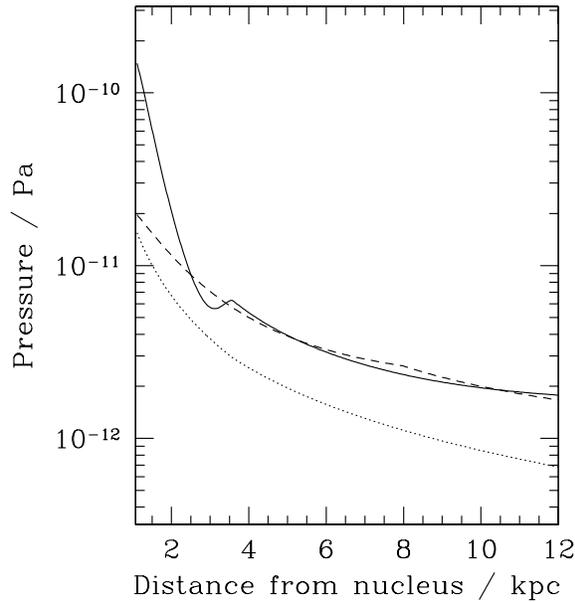}
    \caption{Pressure profiles along the axis of 3C\,31 from a conservation-law
    analysis \citep{LB02b}. The full line is the internal pressure, the dashed
    line the external pressure deduced from {\sl Chandra} observations and the
    dotted line the synchrotron minimum pressure. Note that the jet is
    overpressured between 1 and 3.5\,kpc from the nucleus.\label{fig:pressure}}
  \end{center}
\end{figure}

\begin{figure}
  \begin{center}
    \includegraphics[width=\columnwidth]{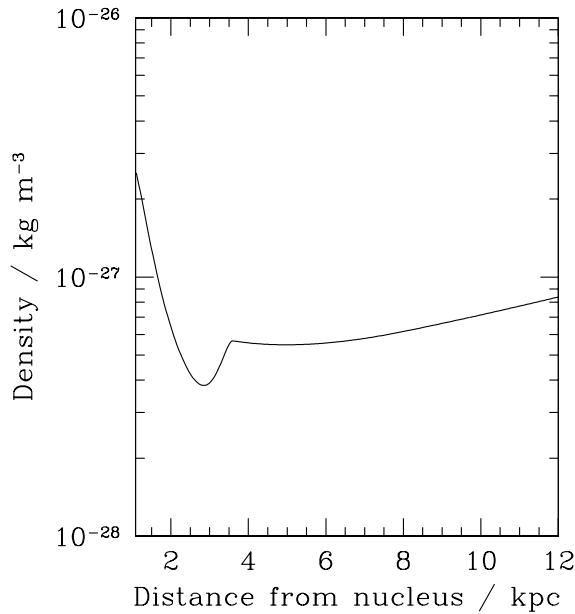}
    \caption{Profile of internal density for 3C\,31, as deduced from the
    conservation-law analysis of \cite{LB02b}.\label{fig:density}}
  \end{center}
\end{figure}

\begin{figure}
  \begin{center}
    \includegraphics[width=\columnwidth]{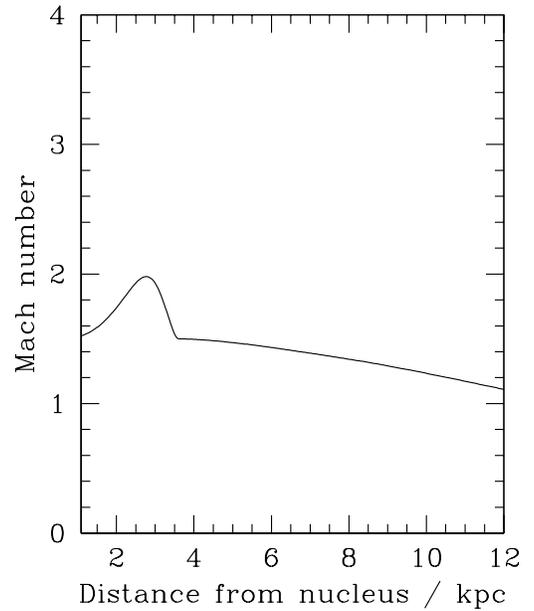}
    \caption{Profile of internal Mach number for 3C\,31, as deduced from the
    conservation-law analysis of \cite{LB02b}.\label{fig:machno}}
  \end{center}
\end{figure}

\begin{figure}
  \begin{center}
    \includegraphics[width=\columnwidth]{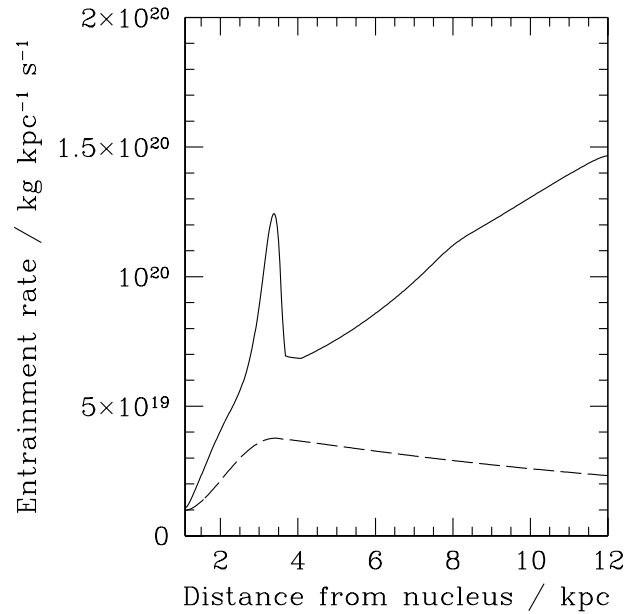}
    \caption{Profiles of entrainment rate for 3C\,31. The full line is deduced
    from the conservation-law analysis and the dashed line is an estimate of
    mass injection due to stellar mass loss \citep{LB02b}.\label{fig:entrain}}
  \end{center}
\end{figure}

For 3C\,31 the key derived parameters at 1\,kpc from the nucleus (the start of
the modelled region) are:
\begin{description}
\item [Mass flux] $\approx$ 0.0005 solar masses / year.
\item [Energy flux] $\approx$ 1.1 $\times$ 10$^{37}$\,W.
\item [Pressure] $\approx$ 1.5 $\times$ 10$^{-10}$\,Pa.
\item [Density] $\approx$ $2 \times 10^{-27}$\,kg\,m$^{-3}$.
\item [Mach number] $\approx$ 1.5.
\item [Entrainment rate] $\approx 1.2 \times 10^{10}$\,kg\,kpc$^{-1}$\,s$^{-1}$.
\end{description}
Profiles of internal and external pressure, density, Mach number and entrainment
rate are shown in Figs~\ref{fig:pressure} -- \ref{fig:entrain}.

\subsection{What are the jets made of?}

Unfortunately, our analysis cannot give a definitive answer to this question,
but they do imply that the jets are extremely light. At the innermost modelled
point, we infer a density of $2.3 \times 10^{-27}$\,kg\,m$^{-3}$, equivalent to
1.4 protons m$^{-3}$. For a power-law energy distribution of radiating
electrons, each accompanied by a proton for charge balance, there must be a
minimum Lorentz factor (the infamous $\gamma_{\rm min}$) in the range 20 -- 50.
Radiation from such low-energy electrons is at frequencies well below those
observable. We cannot exclude any of the following alternatives (or, indeed, a
mixture of them).
\begin{enumerate}
\item Pure electron-positron plasma with an excess of particles over a power-law
  distribution at low energies.
\item Electron-positron plasma with a power-law distribution and a small amount
  of mixed-in thermal plasma.
\item A power-law distribution of electrons, each accompanied by a ``cold''
  (i.e. not ultrarelativistic) proton and $\gamma_{\rm min} \approx$ 20 -- 50.
\end{enumerate}

We have, however, estimated the mass injection into the jet from stellar mass
loss within 1\,kpc of the nucleus. This is commensurate with the mass flux we
infer at that distance, and entirely consistent with the second alternative of
electron-positron plasma and entrained thermal material. We have also compared
the mass injection rate expected from stars with our inferred entrainment rate
on larger scales (Fig.~\ref{fig:entrain}). The two rates are comparable at
1\,kpc, but an additional source of mass -- presumably provided by
boundary-layer entrainment -- is clearly required at large distances, where the
stellar density falls off.

\section{Adiabatic models and flux-freezing}
\label{Adiabatic}

\begin{figure}
  \begin{center}
    \includegraphics[width=7.75cm]{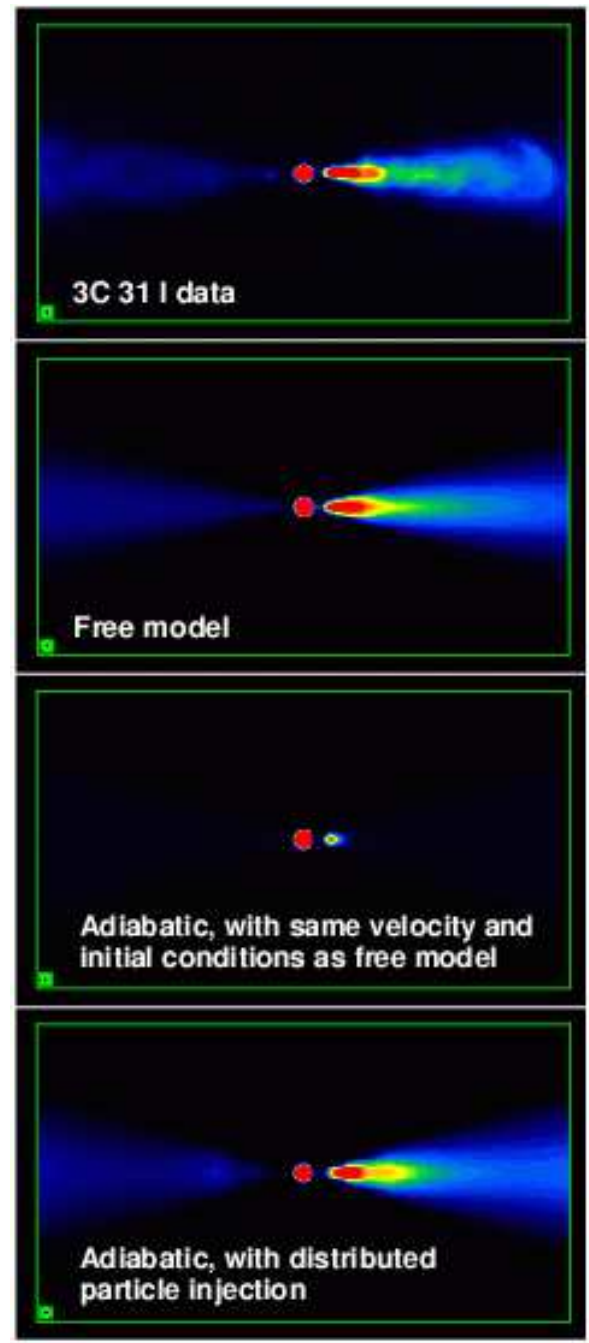}
    \caption{Observed and model brightness distributions for 3C\,31,
    illustrating the effects of adiabatic energy losses. From the top: 
    (a) observed brightness distribution at a resolution of 0.75\,arcsec FWHM;
    (b) free model from \cite{LB02a}: (c) adiabatic model with the same velocity
    field and initial
    conditions (emissivity and field structure) as the free model; (d) adiabatic
    model with distributed particle injection between 2 and 6\,kpc from the
    nucleus. See \cite{LB04} for further details.\label{fig:adiabat}}
  \end{center}
\end{figure}

\begin{figure}
  \begin{center}
    \includegraphics[width=\columnwidth]{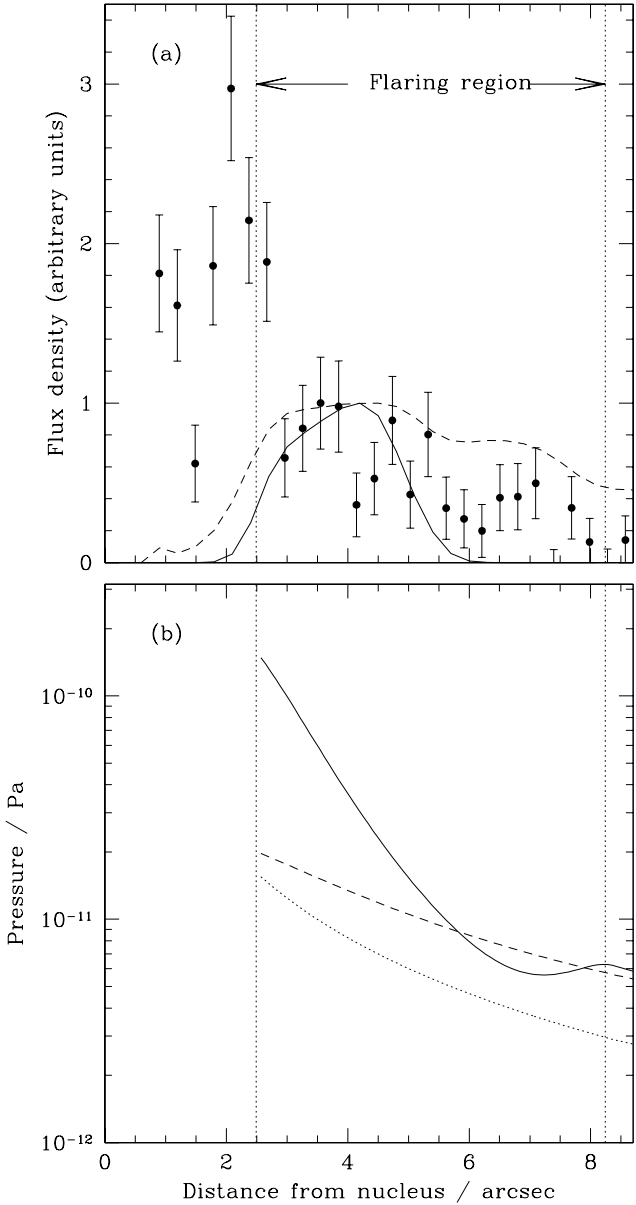}
    \caption{(a) Intensity profiles along the main jet axis for 3C\,31. The
    points show the observed X-ray emission \citep{Hard02}, the full line is the
    particle injection function inferred from our adiabatic models (arbitrarily
    normalised) and the dashed line is the observed radio emission.
    (b) Pressure profiles from Fig.~\ref{fig:pressure}, for comparison. The
    model curves in both panels start at 1\,kpc from the nucleus and do not
    attempt to represent the inner jet. See
    \cite{LB04}.\label{fig:adprofile}}
  \end{center}
\end{figure}

To proceed beyond the purely empirical descriptions of the jets presented in
Sections~\ref{Free} and \ref{Physparms}, we must make further assumptions about
processes that affect the components of the modelled emissivity -- the
relativistic particle energy spectrum and the strength of the magnetic field.
The separation of the emissivity into its components is ill-determined unless
inverse-Compton emission can be detected from the synchrotron-emitting electrons
(not yet the case for any FR\,I jet) and there are, as yet, no prescriptive
theories for dissipative processes such as particle acceleration and field
amplification or reconnection in conditions appropriate to FR\,I radio jets
\citep{DeY04}.

The energy-loss processes for the radiating particles can be quantified,
however.  It is inevitable that the particles will suffer adiabatic losses as
the jets expand and synchrotron and inverse Compton losses are likely to be
negligible by comparison at radio frequencies.
It is therefore worthwhile to compare the
observations with models in which the radiating particles are accelerated before
entering the region of interest and then lose energy only by the adiabatic
mechanism while the magnetic field is frozen into and convected passively with
the flow, which is assumed to be laminar.  Following conventional usage, we
refer to such models as {\em adiabatic}.

Analytical expressions for magnetic-field strength and emissivity were given for
quasi-one-dimensional, non-relativistic jets by \cite{Burch79} and \cite{Fan82}
and extended to the relativistic case by \cite{Bau97}.  These formulae are not
valid if there is velocity shear in a direction perpendicular to any component
of the field. We therefore developed a more general approach which allows the
inclusion of more realistic geometries and velocity fields and calculates the
observed brightness and polarization structure given prescribed initial
conditions on the emissivity and field \citep{LB04}.  We found that adiabatic
models give a fair description of the observed brightness and polarization
distributions in the outer parts of the jets in 3C\,31, but fail in the flaring
region, where they predict a far steeper decline in emissivity than is
observed. This is illustrated in Fig.~\ref{fig:adiabat}, where we show the
brightness distribution predicted by an adiabatic model of 3C\,31, given the
velocity field and initial conditions of the free model and starting at 1\,kpc
from the nucleus. The observed polarization distribution is also inconsistent in
detail with flux-freezing in a laminar flow.

We showed, however, that a modified adiabatic model can still be fitted to the
total intensity of the flaring and outer regions if we add distributed injection
of relativistic particles which then evolve adiabatically; the region where
these particles must be injected is also one where there is independent evidence
for recent particle acceleration from the detection of X-ray synchrotron
radiation \citep{Hard02} and of a local over-pressure in the jet from dynamical
arguments (Fig.~\ref{fig:pressure}).  The predicted brightness distribution is
shown in the bottom panel of Fig.~\ref{fig:adiabat} and the inferred profile of
particle injection is compared with that of the observed X-ray emission in
Fig.~\ref{fig:adprofile}.

\begin{figure}
  \begin{center}
    \includegraphics[width=\columnwidth]{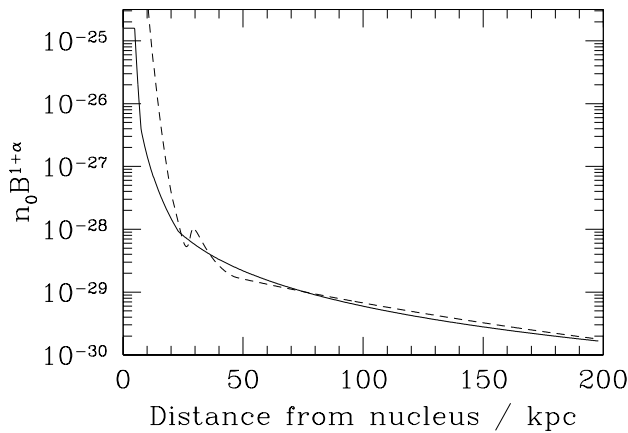}
    \caption{Profiles of $n_0 B^{1+\alpha}$ for 1553+24, from \cite{CL04}. The
    units are as in Fig.~\ref{fig:ngc315emiss}. The full line shows the results
    from the free model and the dashed line the adiabatic model prediction,
    normalised to match at large distances from the nucleus. Note the good
    agreement over a factor of 20 in distance and 50 in
    emissivity.\label{fig:1553profile}}
  \end{center}
\end{figure}

A simpler analysis of 1553+24 shows that the emissivity profile is fit
surprisingly well by the quasi-one-dimensional adiabatic approximation
at large distances from the nucleus (Fig.~\ref{fig:1553profile}). We plan more
sophisticated modelling to determine whether the slow outer regions of FR\,I
jets can be adequately described by the adiabatic approximation.

\section{Conclusions}

We conclude that FR\,I jets are decelerating, relativistic flows, which we can
now model quantitatively. Their three-dimensional distributions of velocity,
emissivity and field ordering can be inferred by fitting to deep, well-resolved
radio images in total intensity and linear polarization. Application of
conservation of mass, energy and momentum allows us to deduce the variation of
density, pressure and entrainment rate along the jets. Boundary-layer entrainment
and mass input from stars are probably both important in slowing the
jets. Adiabatic models and flux freezing do not work everywhere, but do describe
the observations at large distances from the nucleus. Finally, we infer from the
radio observations alone that fresh radiating particle must be injected where
the jets are fast - this is precisely where we detect X-ray synchrotron
emission.

\section*{Acknowledgments}
We thank our collaborators, in particular Martin Hardcastle, Diana Worrall, Bill
Cotton and Paola Parma, for providing both radio and X-ray data and wise advice.
The National Radio Astronomy Observatory is a facility of the National Science
Foundation operated under cooperative agreement by Associated Universities, Inc.

\end{document}